\documentclass [article,epsfig,11pt,floatfix,amsmath,amssym,pre,amssymb,aps,superscriptaddress]{revtex4}

\usepackage[english]{babel}
\usepackage{graphicx}
\usepackage{subfigure}
\usepackage{setspace}
\usepackage{color}
\usepackage{amssymb}
\usepackage{amsmath}
\usepackage{mathtools}

\usepackage{setspace}

\begin{document}
\newcommand{\rv}{{\mathbf r}}
\newcommand{\qv}{{\mathbf q}}
\newcommand{\av}{{\mathbf a}}
\newcommand{\cH}{{\cal H}}
\newcommand{\lB}{\ell_B}
\newcommand{\apa}{a_\parallel}
\newcommand{\ape}{a_\perp}
\newcommand{\Sv}{\hat{{\bf S}}}
\newcommand{\ev}{\hat{{\bf e}}}
\newcommand{\tv}{\hat{{\bf t}}}
\newcommand{\Bf}{\bar{f}}
\renewcommand{\d}[1]{\ensuremath{\operatorname{d}\!{#1}}}

\newcommand{\bra}[1]{\ensuremath{\left\langle#1\right|}}
\newcommand{\ket}[1]{\ensuremath{\left|#1\right\rangle}}
\newcommand{\bracket}[2]{\ensuremath{\left\langle#1 \vphantom{#2}\middle|  #2 \vphantom{#1}\right\rangle}}
\newcommand{\matrixel}[3]{\ensuremath{\left\langle #1 \middle| #2 \middle| #3 \right\rangle}}

\newcommand{\braZ}[1]{\ensuremath{\left\langle#1\right}}
\newcommand{\ketZ}[1]{\ensuremath{\left\right\rangle}}


\title{Statistical mechanics of specular reflections from fluctuating membranes and interfaces}

\date{\today}

\author{Amir Azadi}
\affiliation{Department of Physics, Harvard University, Cambridge, MA 02138, USA}
\affiliation{Wolfram Research, Somerville, MA 02144, USA}

\author{David R. Nelson}
\affiliation{Department of Physics, Harvard University, Cambridge, MA 02138, USA}
\affiliation{School of Engineering and Applied Sciences, Harvard University, Cambridge, MA 02138, USA}

\begin{abstract}
We study the density of specular reflection points in the geometrical optics limit when light scatters off fluctuating interfaces and membranes in thermodynamic equilibrium. We focus on the statistical mechanics of both capillary-gravity interfaces (characterized by a surface tension) and fluid membranes (controlled by a bending rigidity) in thermodynamic equilibrium in two dimensions. Building on work by Berry, Nye, Longuet-Higgins and others, we show that the statistics of specular points is fully characterized by three fundamental length scales, namely, a correlation length $\xi$, a microscopic length scale $\ell$ and the overall size $L$ of the interface or membrane. By combining a scaling analysis with numerical simulations, we confirm the existence of a scaling law for the density of specular reflection points, $n_{spec}$, in two dimensions, given by $n_{spec}\propto\ell^{-1}$ in the limit of thin fluctuating interfaces with the interfacial thickness $\ell\ll\xi_I$. The density of specular reflections thus diverges for fluctuating interfaces in the limit of vanishing thickness and shows no dependance on the interfacial capillary-gravity correlation length $\xi_I$. Although fluid membranes under tension also exhibit a divergence in $n_{spec}\propto\left(\xi_M\ell\right)^{-1/2}$, the number of specular reflections in this case can grow by decreasing the membrane correlation length $\xi_{M}$.
 \end{abstract}

\maketitle

\section{Introduction}
The intricate dancing pattern of bright points and curves which can be seen as sea surface reflections or as caustics at the bottom of a swimming pool, in the presence of sunlight, echoes the geometry of rippling water surfaces. These specular points and caustic reflections were known historically to the ancient Greeks, and were sketched and appreciated by Leonardo da Vinci \cite{vinci}.
Provided light wavelengths are short compared to ripple sizes, they obey the principles of classical geometric optics as embodied in Fermat's principle \cite{nye}. However, systematic studies of the statistics of the specular points and optical caustics which encode the geometrical features of fluctuating surfaces are relatively recent. Among the theoretical and experimental studies \cite{berry, berry_2,long_1,long_2,long_3, ref1, ref2, ref3, ref4}, of particular interest is the pioneering work by Munk and Cox \cite{Munk} who developed a method to measure the roughness of the sea surface agitated with wind driven waves, thus determining the effect of wind on the mean square surface slope from the spatial distribution of specular reflections. 
  
The distribution of specular reflections is also connected to problems in condensed matter and stochastic processes which include the statistics of zero crossings of random functions, and the noise currents embodied in the Shot noise \cite{rice}.
The statistics of maxima, minima and saddle points, for two dimensional membranes and interfaces is related to the density of specular reflection points of a stochastic scalar field, in one and two
dimensions. The generalization to a variety of physical systems is the subject of extensive works by Berry, Upstill  \cite{berry,berry_2}, Longuet-Higgins \cite{long_1,long_2,long_3} and others inspired in part by the catastrophe theory of caustic formation by light \cite{nye}. In Refs. \cite{lax, halp_2}, Halperin {\it et al.}, have studied related statistical properties of the zeros of the $n$-dimensional vector fields in the $d$-dimensional space. These points for $n=d$ (or curves in $n=d-1$), characterize the topological singularities in the orientation of a vector field \cite{min}. 
Here we apply related ideas to specular reflections of various types membranes and interfaces in thermodynamic equilibrium. By ``interface" we mean a boundary between a liquid and gas phase, where the restoring forces are gravity and surface tension. We use ``membrane" to denote the system such as lipid bilayers with aqueous phase above and below, where the dominant forces are a bending rigidity and an effective surface tension. 

An analysis of specular reflections off an undulating water surface illuminated by a distant light source such as the sun \cite{long_1,long_2,long_3} presents fascinating problems linked to the rich statistical dynamics of capillary and gravity waves. With wind-driven wave excitations, generated by the non-equilibrium dynamics of the atmosphere, one must deal with complex nonlinear mode-couplings and energy transfers across length scales associated with driven wave turbulence \cite{a,b,c}. In this paper, we investigate two simpler models of reflecting surfaces, associated with interfaces and membranes in thermodynamic equilibrium in two dimensions. Exact results are readily generated, which can be checked by straightforward computer simulations. In this sense, our investigation is analogous to the ``absolute equilibrium" models of homogeneous, isotropic turbulence, which are interesting in their own right and can sometimes provide insights into more complex problems, such as the direction of turbulent energy cascades \cite{d, e}.

\begin{figure*}
\center\includegraphics[width=0.9\textwidth]{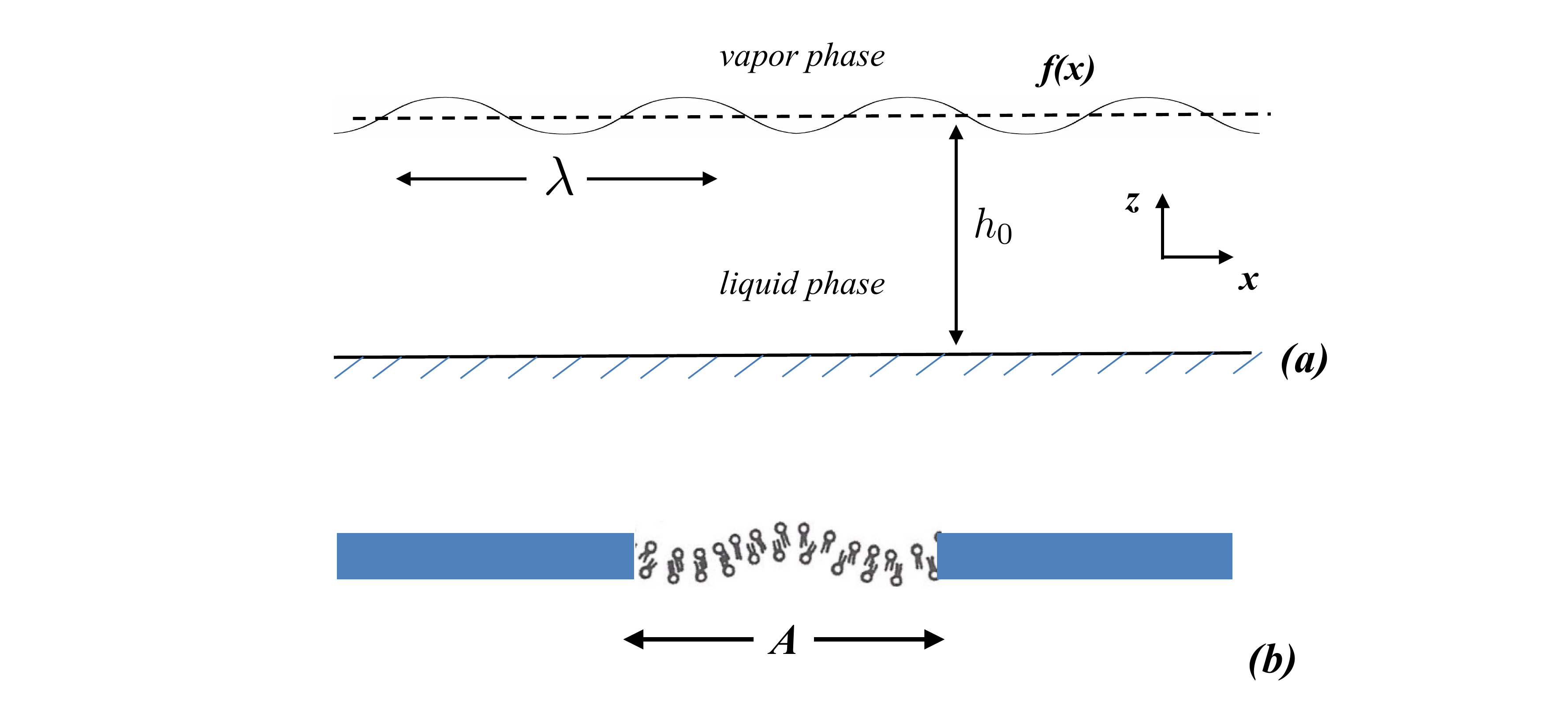}
\caption{(a) An interface undulation with wavelength $\lambda$ at the liquid-vapor interface of average height $h_{0}$. The dominant restoring force is gravity or surface tension, depending on whether $\lambda$ is greater than or less than the interfacial capillary-gravity length scale $\xi_I=\sqrt{\sigma/\rho_{0}g}$. (b) A lipid bilayer membrane in water (with equal osmotic pressures above and below the barrier), is suspended across a hole with area $A$. Membrane undulations in this case are controlled by a bending rigidity $\kappa$ and by the constraint that the membrane have a fixed projected area $A$, which acts like an effective surface tension $\alpha$. The result is a membrane correlation length $\xi_M=\sqrt{\kappa/\alpha}$.}
\label{fig0}
\end{figure*}

Our first system, designed to model specular reflections off, say, air-water interfaces, assumes that the configurations $f(\vec{x})$, of a single-valued interface height profile in $d$-dimensions (see Fig. (\ref{fig0}(a)) are governed by an equilibrium probability distribution
$P_{int}\left[f(\vec{x})\right]\propto\exp\left[-F_{int}[f(\vec{x})]\right/k_{B}T]$, with an equilibrium interfacial free energy given by 
\begin{eqnarray}
\label{Fe_intro}
F_{int}\left[f(\vec{x})\right]=\frac{1}{2}\int d^{d-1}x \left[\sigma|\vec{\nabla} f(\vec{x})|^2+\rho_0 g f^2(\vec{x})\right].
\end{eqnarray}		
Here, the first term arises from the gradient expansion of the contribution from a surface tension $\sigma$, 
$F_{s}=\sigma\int d^{d-1} x \sqrt{1+ |\vec{\nabla} f(\vec{x})|^2}$ and the second from the gravitational potential energy, $\rho_0g\int_{0}^{h(\vec{x})}f'df'=\frac{1}{2}\rho_0 g h^2(\vec{x})$, integrated over the surface of an incompressible liquid with height $h(\vec{x})$ and mass density $\rho_0$, in equilibrium with a vapor phase of negligible density. Here $g$ is the gravitational constant and $f(\vec{x})$ is the deviation of the fluid height from its equilibrium value $h_0$. The interfacial length scale $\xi_I=\sqrt{\sigma/\rho_0 g}$ marks the boundary between capillary and gravity wave excitations \cite{f}. A similar long-wavelength description describes solid-vapor interfaces above the roughening transition \cite{g}.

Our second model describes specular reflections off a membrane, e.g., a thermally fluctuating lipid bilayer, suspended in water across a hole of fixed cross-sectional area with zero osmotic pressure difference (see Fig. (\ref{fig0}(b)). We use the Monge representation to describe a nearly flat membrane embeded in $d$ dimensions located at $\vec{r}\left(x_1,...,x_{d-1}\right)=\left[x_1,...,x_{d-1}, f(\vec{x})\right]$. In the absence of this constraint, fluctuating membranes are controlled by a bending free energy, $F_{b}=\frac{1}{2}\kappa\int d^{d-1}x |\nabla^2 f(\vec{x})|^2$, characterized by a bending rigidity $\kappa$. We assume an incompressible lipid bilayer, and implement the constraint of fixed total membrane area, $A_0$ via a Lagrange multiplier $\alpha$, and thus describe the statistical mechanics by a total free energy, $F_{tot}=F_{b}-\alpha\int d^{d-1} x\sqrt{1+|\vec{\nabla} f(\vec{x})|^2}$. Upon expanding the square root in this integral and neglecting a constant term, the probability of a particular membrane configuration is given by 
$P_{mem}\left[f(\vec{x})\right]\propto\exp\left[-F_{mem}[f(\vec{x})]/k_{B}T\right]$, where 
\begin{eqnarray}
\label{Fe_intro_mem}
F_{mem}\left[f(\vec{x})\right]=\frac{1}{2}\int d^{d-1}x \left[\kappa|\nabla^2 f(\vec{x})|^2+\alpha|\vec{\nabla} f(\vec{x})|^2\right].
\end{eqnarray}	
The constant $\alpha$ is fixed by the condition that 
\begin{eqnarray}
\label{l_const}
\int d^{d}x\left\langle\sqrt{1+|\vec{\nabla} f(\vec{x})|^2}\right\rangle=A_0>A,
\end{eqnarray}	
where the brackets indicate a $\alpha$-dependent thermal average over $P_{mem}\left[f(\vec{x})\right]$ in an ensemble specified by Eq. \ref{Fe_intro_mem}. Here $A_{0}$ is the area an incompressible membrane would have at zero temperature and $A$ is the projected area, shown in Fig. 1b. 
In the small gradient approximation, assumed throughout this paper, $\alpha$ is thus determined by,
$\int d^{d-1}x\left\langle |\vec{\nabla} f(\vec{x})|^2\right\rangle=A_{0}-A$
(If $A<A_0$, $\alpha$ is negative and the membrane can be subject to a buckling instability; in this paper, we assume $\alpha>0$). Note that the effect of the constraint of fixed area leads to a tension-like contribution to the effective free energy. Eq. (\ref{Fe_intro_mem}) now defines a membrane length scale $\xi_{M}=\sqrt{\kappa/\alpha}$. The dominant restoring force for fluctuations with wavelengths $\lambda$ less than $\xi_{M}$ is the bending rigidity, while the tension term dominates for $\lambda>\xi_{M}$. Similar free energies arise for one-dimensional polymers in the wormlike-chain approximation with a pulling force applied to the ends \cite{h}.

In this article, by combining analytics and numerical simulations, we study the interplay between the geometry and statistical mechanics of specular reflection with material properties that describe membranes and interfaces in thermal equilibrium.
We begin by calculating the number of zeros of a random Gaussian scalar field, $f(x)$ (or more generally the number of times $f(x)=y$, some fixed height), controlled by probability distributions such as Eqs. (\ref{Fe_intro}) and (\ref{Fe_intro_mem}) using path integral methods. A generalization of this method allows us to calculate the distribution of specular points, generated by a light source far from the plane of the membrane or interface, as seen by a distant observer, due to reflections from a random surface. 

In this paper, we focus for simplicity on $d=2$, {\it i.e.} one-dimensional membranes and interfaces. We hope to publish a paper on specular reflections off two-dimensional surfaces in $d=3$, including an experimental test using reflecting undulating surfaces generated by a three-dimensional printer, in the future.  

In the following sections, we focus first on the density of specular points for the special case where the light source and the observer both reside on a single line perpendicular to the average plane of the reflecting surface. We then adapt this calculation, using the paraxial approximation \cite{nye}, to the case of grazing angles, directions far from the average surface normal where the observer is nevertheless remains far from the reflecting interfaces or membranes.
These calculations allow us to explore how the density of specular points is related to the correlation lengths, $\xi_I$ and $\xi_M$, introduced above to characterize the height-height correlation functions of membranes and interfaces. 
After treating membranes and interfaces in two dimensions, we explore certain geometrical features of the fluctuating surfaces and membranes such as average mean curvature, which is closely related to the density of specular reflection points. The interface and membrane correlation lengths $\xi_I$ and $\xi_M$ determine the density of specular reflection points, when combined with a microscopic length scale $\ell$ and an overall interface or membrane size $L$. Having tabulated the results of specular reflection theory, we then test the validity of the scaling laws for the density of specular points with Monte-Carlo simulations of interfaces and membranes with varying correlation lengths.

\section{Equilibrium thermodynamics of fluctuating interfaces and membranes}
\subsection{Free energy and height correlations in two dimensions}
The configuration of a one-dimensional fluctuating interface or membrane (curve) in two dimensions can be described by the height function $f(x)$, which we assume is single valued, that characterizes the deviation from the flat state. A long wavelength free energy that can be used to describe both interfaces and membranes is given by, 
\begin{eqnarray}
\label{Fe}
F_{1d}&=&\frac{1}{2}\int dx \left[c(\nabla^2 f)^2+b\left(\vec{\nabla} f\right)^2+a f^2\right].
\end{eqnarray}
If we set $c=0$, and take $a$ to be proportional to the gravitational constant and $b$ to the surface tension, we obtain the capillary-gravity interface model of Eq. (\ref{Fe_intro}).
Note that the first term in eq. (\ref{Fe}) $\sim c(\nabla^2 f)^2$, could then be introduced in the capillary-gravity model to suppresses short wave length fluctuations. Here we shall instead simply forbid wavevectors with $q\gtrsim\pi/\ell$ in the Fourier expansion of the free energy,
\begin{eqnarray}
\label{Fe_ab}
F_{I}&=&\frac{1}{2}\int dx \left[b\left(\vec{\nabla} f\right)^2+a f^2\right],
\end{eqnarray}
where $\ell$ is a microscopic length of order the interface thickness.  

On the other hand, neutrally buoyant membranes subject to an isotropic tension due to an area constraints, as in Fig. 1, can be described by setting $a=0$,
\begin{eqnarray}
\label{Fe_bc}
F_{M}&=&\frac{1}{2}\int dx \left[c(\nabla^2 f)^2+b\left(\vec{\nabla} f\right)^2\right].
\end{eqnarray}  
The probability of a given equilibrium configuration $f(x)$ at absolute temperature $T$ is given by $P(f)\propto\exp\left[-F/k_{B}T\right]$, where $k_{B}$ is Boltzmann's constant. Henceforth, we rescale energy units such that $k_{B}T=1$. It will be convenient to expand $f(x)$ in Fourier modes,
$f(x)=\frac{1}{L}\sum_{q}f_{q}e^{iqx}$, where $L$ is the macroscopic interface or membrane size, and we assume periodic boundary conditions. 

We first calculate the 
height correlation function which describes fluctuations of interfaces and membranes in thermodynamic equilibrium.  
Upon passing to Fourier space, the correlation function for the out-of-plane fluctuation of the two dimensional interface is described, in the limit of small $c$, by 
\begin{eqnarray}
\label{Cc}
C(y)&=&\langle f(x_{0})f(x_{0}+y)\rangle
\approx\frac{1}{2\sqrt{ab}}e^{-|y|/\xi_I}
\end{eqnarray}
where the capillary-gravity interfacial correlation length is $\xi_I=\sqrt{b/a}$. This limit sends the effective ultraviolet cutoff in Fourier space $1/\ell=\sqrt{b/c}$, to large values, and allows us to use the contour integration for $q$ in the complex plane with $\delta=b^{2}-4ac>0$ (see Appendix A for details).
Similar methods show that, in contrast, the second derivative of this correlation function depends explicitly on the ultraviolet cutoff parameter $c$. When $c$ becomes small, we have
\begin{eqnarray}
\label{Cc_d}
C^{(2)}(y)
\approx-\frac{1}{2}\frac{1}{\sqrt{bc}}e^{-|y|/\xi_I},
\end{eqnarray}
where $C^{(2)}(y)\equiv\frac{d^2C(y)}{dy^2}$. To determine quantities such as the densities of zero crossings and specular points, we shall need (see below) the ratio of the correlation function to its second derivative at $y=0$. This quantity is governed by two distinct length scales,
\begin{eqnarray}
\label{Ccappc}
\frac{C^{(2)}(0)}{C(0)}&\approx&-\sqrt{\frac{a}{c}}=-\frac{1}{\ell\xi_I}
\end{eqnarray}
where the last inequality holds, provided we regard the free energy Eq. \ref{Fe} with $a$, $b$ and $c>0$ as model of an interface with an ultraviolet cutoff imposed by the parameter $c$ and 
where the effective short distance cutoff for this model is $\ell=\sqrt{c/b}$. As discussed in Appendix A, we work in a regime such that $\ell\ll\xi_M\ll L$, where $L$ is the macroscopic size of the interface. The first inequality is equivalent to the condition $\delta=b^2-4ac\gg0$.

\subsection{Capillary-gravity wave and fluctuating membrane models} We now focus on the two limiting cases $a=0$ and $c=0$ in more detail. The free energy given by Eq. (\ref{Fe}) with finite $a$ and $b$ and vanishing $c$ is a model of capillary-gravity waves at the liquid-air interface, as discussed above. In this case, $a$ is proportional to the liquid density $\rho$ and the gravitational constant $g$, $a\propto\rho g$, and $b$ describes the line tension at the one dimensional interface. It is instructive to 
recompute the correlation functions for the capillary-gravity model in the Fourier space with $c=0$, while imposing a hard upper cutoff $\pi/\ell$ on the allowed wavevectors, 
\begin{subequations}
\begin{align}
\label{Co2_1}
C(0)&=\frac{1}{2\pi}\int^{+\pi/\ell}_{-\pi/\ell}\frac{dq}{a+b q^2}=\frac{\xi_I}{b}\arctan\left[\frac{\xi_I\pi}{\ell}\right]\approx\frac{1}{\sqrt{ab}}=\frac{\xi_I}{b}
\\
\nonumber\\
\label{Co2_2}
C^{(2)}(0)&=\frac{1}{2\pi}\int^{+\pi/\ell}_{-\pi/\ell}\frac{-q^2 dq}{a+b q^2}\approx-\frac{1}{b\ell}
\\
\nonumber\\
\label{Co2_3}
C^{(4)}(0)&=\frac{1}{2\pi}\int^{+\pi/\ell}_{-\pi/\ell}\frac{q^4 dq}{a+b q^2}\approx\frac{\pi^2}{3b\ell^3}\end{align}
\end{subequations}
where $\ell$, the interface thickness, sets the ultraviolet cutoff, and the capillary-gravity wave interfacial correlation length is $\xi_I=\sqrt{b/a}$. In this hydrodynamic treatment, we again consider the regime $\ell\ll\xi_I\ll L$. Note that this alternative ultraviolet cutoff leads to $C^{(2)}(0)/C(0)=-1/\ell\xi_I$, a result in agreement with Eq. (\ref{Ccappc}), where the short distance was imposed by $c$. Here, $C^{(4)}(0)=\lim_{y\rightarrow 0} C^{(4)}(y)=\lim_{y\rightarrow 0}\frac{d^{4}C(y)}{dy^4}=\lim_{y\rightarrow 0}\langle f''(x+y)f''(x)\rangle$, we shall need it in later sections. 

We now discuss the case $a=0$, i.e. a membrane with a tension in two dimensions, or equivalently a semi-flexible polymer in two dimensions with a force applied to the ends \cite{h}.
In this case, $c$ represents the bending rigidity and $b$, a force or tension.
The height correlation function for a $1$-dimensional membrane (or stretched semi-flexible polymer) with a tension in two dimensions at zero separation is given by,
\begin{subequations}
\begin{align}
\label{Ccapp}
C(0)&=\frac{1}{\pi}\int^{\infty}_{\pi/L}\frac{dq}{b q^2+c q^4}\approx\frac{L\xi_{M}}{\sqrt{bc}}\\
\label{Ccapp-b}
C^{(2)}(0)&=\frac{d^2C(y)}{dy^2}\bigg|_{y=0}=\frac{-1}{\pi}\int^{\infty}_{\pi/L}\frac{dq}{b +c q^2}\approx-\frac{1}{2}\sqrt{\frac{1}{bc}}
\\
\label{Ccapp-c}
C^{(4)}(0)&=\frac{d^4C(y)}{dy^4}\bigg|_{y=0}=\frac{1}{\pi}\int^{\pi/\ell}_{\pi/L}\frac{q^2 dq}{b +c q^2}\approx\frac{1}{\ell c}
\end{align}
\end{subequations}
where we impose an infrared cutoff by forbidding wave vectors such that $|q|<q_{min}=\pi/L$ where $L$ is the system size.
We have also defined the membrane correlation length to be, $\xi_{M}=\sqrt{c/b}$. In our long wavelength expansion, we consider the limit of large system size compared to the membrane correlation length. We also assume a short distance cutoff $\ell$, of order the membrane thickness, such that $\ell\ll\xi_{M}\ll L$. Note that the fluctuation amplitude,
$\langle f^2(x)\rangle=C(0)$, diverges with the system's size $L$, but $L$ drops out of Eq. (\ref{Ccapp-b}) in this limit.

\section{Zero crossings of interfaces and membranes in two dimensions}
Before determining the density of specular points for interfaces and membranes, we first adapt ideas of O. Rice \cite{rice} to determine the density of zero crossings, a simpler problem illustrating functional integral techniques used elsewhere in this paper.  
Consider a single-valued random function $f(x)$, which could be the two-dimensional profile of an interface or membrane which can assume the height $z=f(x)$ at various points along the x-axis. We assume that the probability of different configurations is Gaussian and described by a hydrodynamic free energy such as 
Eq. (\ref{Fe}).
We again Fourier analyze the single valued height function $f(x)$, into plane waves:
\begin{eqnarray}
\label{fx_four}
f(x)=\sum_{q} f_{q}e^{iqx}.
\end{eqnarray}
Upon assuming periodic boundary conditions, $f(x+L)=f(x)$, the allowed wave vectors in Eq. (\ref{fx_four}) are, $q_{m}=\frac{2\pi}{L}m$ for $m=0,\pm1,\pm2,...$ . In the limit $L\rightarrow\infty$, we again replace the discrete sum, $\sum_{q_m}$ by an integral over $q$ with $dq=2\pi/L$. 
The requirement that $f(x)$ be real leads to a condition on the complex Fourier amplitudes, namely $f^*_{q}=f_{-q}$ with a Gaussian probability distribution function controlled by the negative exponential of the underlying free energy divided by $k_B T$.
The statistical properties of $f(x)$ are fully controlled by the autocorrelation function $C(x)=\langle f(y)f(x+y) \rangle$, which can be expressed as $C(x)=\int dq E(q)\exp(iqx)$,
where $E(q)=\langle |f(q)|^2\rangle$ is the ``power spectrum" in $q$ space.

Consider the joint probability distribution $p(f=y,f_{1};x)$ such that $p(f=y,f_{1};x) df$ gives the chance that $f(x)$ in the small interval $(x+dx)$ takes a value in the range $f\in [y,y+df]$ with an undetermined slope $f_{1}=df/dx$. We are interested in the probability that $f(x)$ passes through zero, $f=0$ (or in general $f(x)=y$), for any value of the slope $df/dx$ in $[x,x+dx]$. This quantity is given by $\int^{\infty}_{-\infty} df_{1}   p(f=y,f_{1};x) df=dx\int^{\infty}_{-\infty} p(f=y,f_{1};x)|f_{1}|  df_{1}$.
The probability density, $p(f=y,f_{1};x)$ can be expressed as a functional integral in terms of a weighted average on all possible configurations of
$f(x)$,
 \begin{eqnarray}
\label{pd_1}
p\left[f(x)=g,f'(x)=g_1;x\right]&=&\frac{\int \mathcal{D}f(y)\delta\left[f(x)-g\right]\delta\left[f'(x)-g_1\right]e^{-F\left[f(y)\right]}}{\int \mathcal{D}f(y) e^{-F\left[f(y)\right]}}\nonumber\\
\nonumber\\&=&\frac{1}{2\pi \left[-C(0)C^{(2)}(0)\right]^{1/2}}\exp\left[-\frac{f^{2}}{2C(0)}+\frac{f^{2}_{1}}{2C^{(2)}(0)}\right],
\end{eqnarray}
where we define,
\begin{subequations}
\begin{align}
\label{Cx}
C(x)&=\langle f(y)f(x+y) \rangle\\
C^{(2)}(x)&=\frac{d^2 C(x)}{dx^2}=-\langle\frac{df(y)}{dy}\frac{df(x+y)}{dy}\rangle
\end{align}
\end{subequations}
The derrivation of Eq. (\ref{pd_1}) is reported in Appendix B.

We can now calculate the average number of times the function $f(x)$ in the interval $-L/2<x<+L/2$ passes through a specific value $f=y$ (subject to periodic boundary conditions), 
\begin{eqnarray}
\label{N0}
N_{f=y}&=&\int^{+L/2} _{-L/2}dx\int^{\infty}_{-\infty} p(f=y,f_1;x) |f_{1}|  d f_{1}\nonumber\\
&=&\int^{+L/2} _{-L/2}dx\left[\frac{-C^{(2)}(0)}{\pi^2 C(0)}\right]^{1/2}\exp\left[-\frac{f^{2}}{2C(0)}\right]\nonumber\\
&=&L\left[\frac{-C^{(2)}(0)}{\pi^2C(0)}\right]^{1/2}\exp\left[\frac{-f^2}{2C(0)}\right]
\end{eqnarray}
Thus, the number density $n_{f=y}=N_{f=y}/L$ of crossings is given by, 
\begin{eqnarray}
\label{nfy}
n_{f=y}=\left[\frac{-C^{(2)}(0)}{\pi^2 C(0)}\right]^{1/2}\exp\left[-y^{2}/2C(0)\right]
\end{eqnarray}
\begin{figure*}
\center\includegraphics[width=0.9\textwidth]{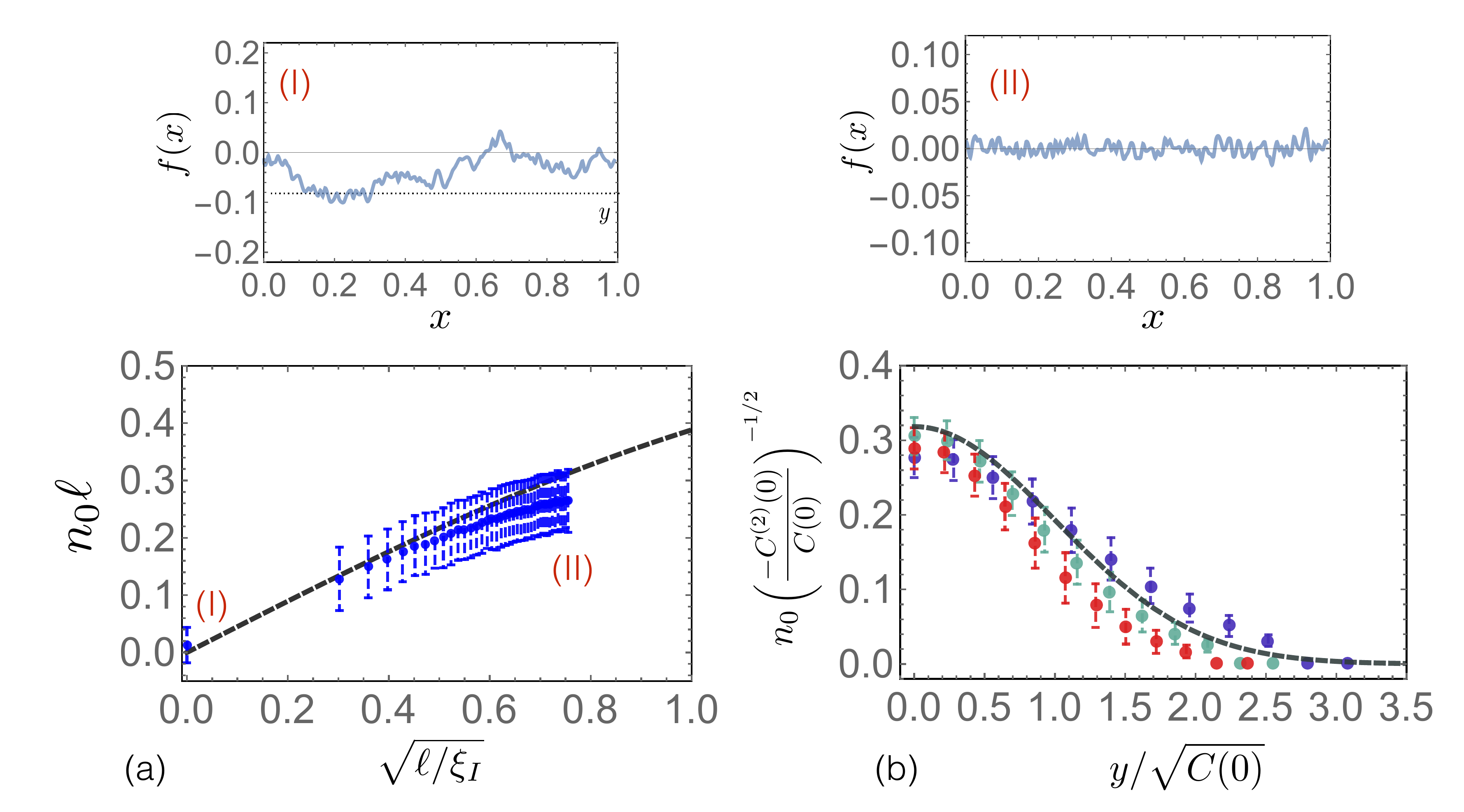}
\label{f1}
\caption{Graphs (I) and (II) (top) depict two sample simulated configurations of fluctuating interfaces for $\sqrt{\ell/\xi_I}=0$ (i.e.we set $a=0$) and $0.9$ respectively. Note the much smaller fluctuations in (II).
(a) The result of the Monte-Carlo simulations for the scaled average density of zero crossings, $n_{0}\ell$, as a function of $\sqrt{\ell/\xi_I}$ for a fluctuating interface in two dimensions approximated by a mesh with $N=120$ sites and $k_{B}T/(b\ell)=0.1$ (see section IVc for details). (b) The normalized expected number of $f(x)=y$ passages for correlation lengths such that $\sqrt{\ell/\xi_I}=0,0.5,0.9$, with blue, green and red data points respectively}
\end{figure*}
\begin{figure*}
\center\includegraphics[width=0.9\textwidth]{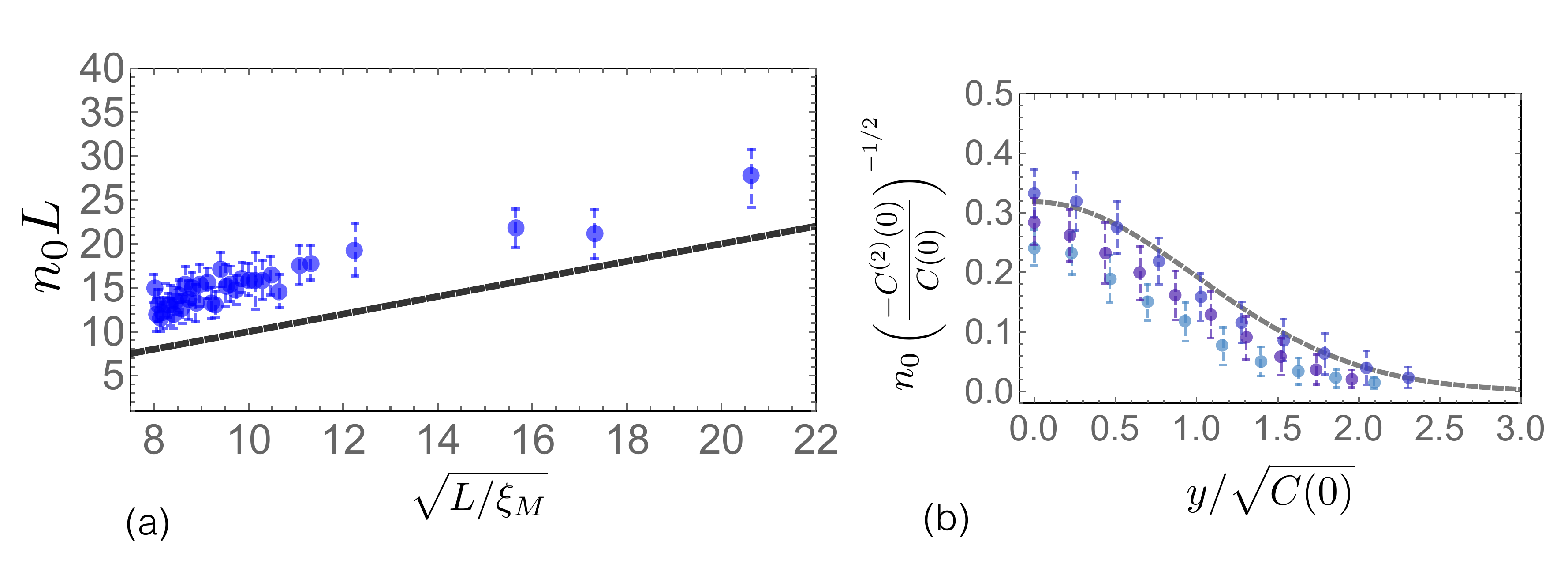}
\label{f_2}
\caption{(a) The result of the simulation for the scaled average density of zero crossings, $n_{0}L$, as a function of $\sqrt{L/\xi_M}$ for a 1D fluctuating membrane with $N=300$ and $k_{B}T/(b\ell)=0.1$. (b) Shows the normalized expected number of $f(x)=y$ passages for series of inverse square of the correlation lengths $\sqrt{L/\xi_M}=8,10,20.6$ which are represented with light blue, dark blue and purple}
\end{figure*}
In order to find the density of zero crossings, $n_{0}$, we need to simply substitute $y=0$ in Eq. (\ref{nfy}).

We can now evaluate the density of crossings for interfaces and membranes from the correlation functions tabulated in Section II.
We can express the number of times per unit length the function $f(x)$ crosses the value $y$ in terms of the correlation lengths $\xi_I$ or $\xi_M$ and the microscopic system thickness $\ell$ and the macroscopic system size $L$, 
\begin{eqnarray}
\label{nxfC}
n_{I}&\approx&\frac{1}{\pi \sqrt{\ell\xi_I}} \exp\left[-\frac{y^{2}}{2\langle f^2(x)\rangle}\right] \quad \textrm{        capillary-gravity wave interface}\\
\nonumber\\
\label{nxfC_2}
n_{M}&\approx&\frac{1}{\pi \sqrt{L\xi_{M}}}  \exp\left[-\frac{y^{2}}{2\langle f^2(x)\rangle}\right]  \quad \textrm{        membrane under tension}
\end{eqnarray}
where $\langle f^2(x)\rangle=C(0)$ is given by Eqs. (\ref{Co2_1}) or (\ref{Ccapp}). Note that in the capillary-gravity interface model the geometric mean of $\xi_I$ and the microscopic cutoff $\ell$ controls the density of zero crossings, whereas in the membrane model the geometric mean of $\xi_{M}$ and the macroscopic system size $L$ is the controlling factor. Thus there are many fewer zero crossings in the softer membrane where longer wavelength fluctuations dominate.  In both cases the number of crossings falls off rapidly where $y\gtrsim\sqrt{\langle f^2(x)\rangle}$. Reasonable agreement with these predictions is found using Monte-Carlo simulations methods (see Sec. IV-C) in Figs 2 and 3 for interfaces and membranes, respectively. 
 
\section{Statistical geometry of specular points on interfaces and membranes in the paraxial approximation}
In this section we use the approach sketched in the last section to determine the distribution of specular reflection points from thermally excited interfaces and membranes in two dimensions. We use simple ideas from the theory of focal singularities \cite{nye} to determine
the geometric condition for the formation of specular points.
In the following two subsections, we determine the density of specular points for two dimensional interfaces and membranes in two limits: large and small angles of incidence. 
\subsection{Nearly normal angles of incidence and reflection}
We assume that the thermally fluctuating boundary, described by Eq. (\ref{fx_four}), is a reflecting surface. A source of light emits wavelengths short compared to the undulations of the interface or membrane, hence we can use geometrical optics. The source resides above the surface, in the $x-z$ plane, at $S=(x_1,z_1)$ and we have the observer at $O=(x_2,z_2)$. Provided  multiple reflections can be neglected, the condition for an arbitrary point, $P=(x,f(x))$ on the reflecting surface $f(x)$, to be a specular point follows from a straightforward application of Fermat's principle \cite{nye}.
The ray from the source $S$ needs to reach to the point $P$ on the surface (see Fig.5.d) and then travel to the observer at $O$. The distances, $l_{1}=OP$ and $l_2=SP$ are given by,
\begin{eqnarray}
\label{li}
l_{i}=\sqrt{(x-x_i)^{2}+(f(x)-z_i)^{2}}\approx z_i-f(x)+\frac{(x_{i}-x)^{2}}{2z_i}
\end{eqnarray}
which holds for $i=1,2$ and where we have taken the limit of positive $z_{i}>>f$, known as paraxial approximation \cite{nye}, in which both the observer and source reside far above a fluctuating surface with small deviations from flatness for an interface or membrane such that $\langle f(x)\rangle=0$. We assume first that both source and observer have $x_{i}\approx0$ as in Fig. 4a, i.e., they lie approximately above the origin of our coordinate system, although possibly with different heights $z_1\neq z_2$. In the next section we consider a more general case with both $x_{i}\ne 0$. Upon assuming a constant propagation velocity in the region above the surface, the path of least time is given by \cite{nye},
\begin{eqnarray}
\label{ltime}
\frac{d}{dx}\left[l_{1}(x)+l_{2}(x)\right]=0,
\end{eqnarray}
which leads via Eq. (\ref{li}) to,
\begin{eqnarray}
\label{ray}
\frac{1}{x}\frac{ d f(x)}{dx}&=&\frac{1}{2}\left[\frac{1}{z_1}+\frac{1}{z_2}\right]\equiv K_1.
\end{eqnarray}
To determine the density of points satisfying this condition for fixed $z_1$ and $z_2$, it is useful to introduce two auxiliary functions, 
\begin{subequations}
\begin{align}
\label{aux_1}
g_{1}(x)&=\frac{1}{x}f'(x)=\frac{1}{x}\frac{ d f(x)}{dx},
\\
\label{aux_11}
 g_{2}(x)&=f''(x)=\frac{d^2 f(x)}{dx^2},
\end{align}
\end{subequations}
and to define $K_1\equiv\frac{1}{2}\left[\frac{1}{z_1}+\frac{1}{z_2}\right]$. Similar to the last section we seek from Eq. (\ref{ray}) the probability of a specular point with $g_{1}=K_1$, somewhere in the interval $[x,x+dx]$,
\begin{eqnarray}
\label{pp1}
p_{spec}(x,K_1)&=&
\int^{\infty}_{-\infty} dg_2 P(g_1=K_1,g_{2};x)|g_2-K_1|\frac{dx}{x}, 
\end{eqnarray}
where $P(g_1,g_2)dg_1dg_2$ gives the probability that the functions in Eq. (\ref{aux_1}) and (\ref{aux_11}) assume the specific values $g_1$ and $g_2$ in the intervals $[g_1,g_1+dg_1]$ and  $[g_2,g_2+dg_2]$. 
To evaluate Eq. (\ref{pp1}), we used the relation $|dg_1|=|\frac{dg_1}{dx}|dx=\frac{|g_1-g_2|}{|x|}dx$. 
For the quadratic energy functionals used here, the probability distribution of $g_1=\frac{1}{x}\frac{df}{dx}$ and the curvature, $g_2$ can be determined by the methods of appendix B to be
 \begin{eqnarray}
\label{dist}
P\left(g_1,g_2;x\right)=(2\pi)^{-1}|M|^{1/2}\exp\left[-\frac{1}{2} \sum_{i,j=1,2}M_{ij}g_{i}g_{j}\right],
\end{eqnarray}
where the $2\times2$ matrix $M_{ij}$ is given by,
\begin{eqnarray}
\label{moment}
M_{ij}=
\begin{bmatrix}
\frac{-x^2}{C^{(2)}(0)}       &  0         \\
         &   \\
      0       &    \frac{1}{C^{(4)}(0)}	
\end{bmatrix}
\end{eqnarray}
\begin{figure*}
\center\includegraphics[width=0.85\textwidth]{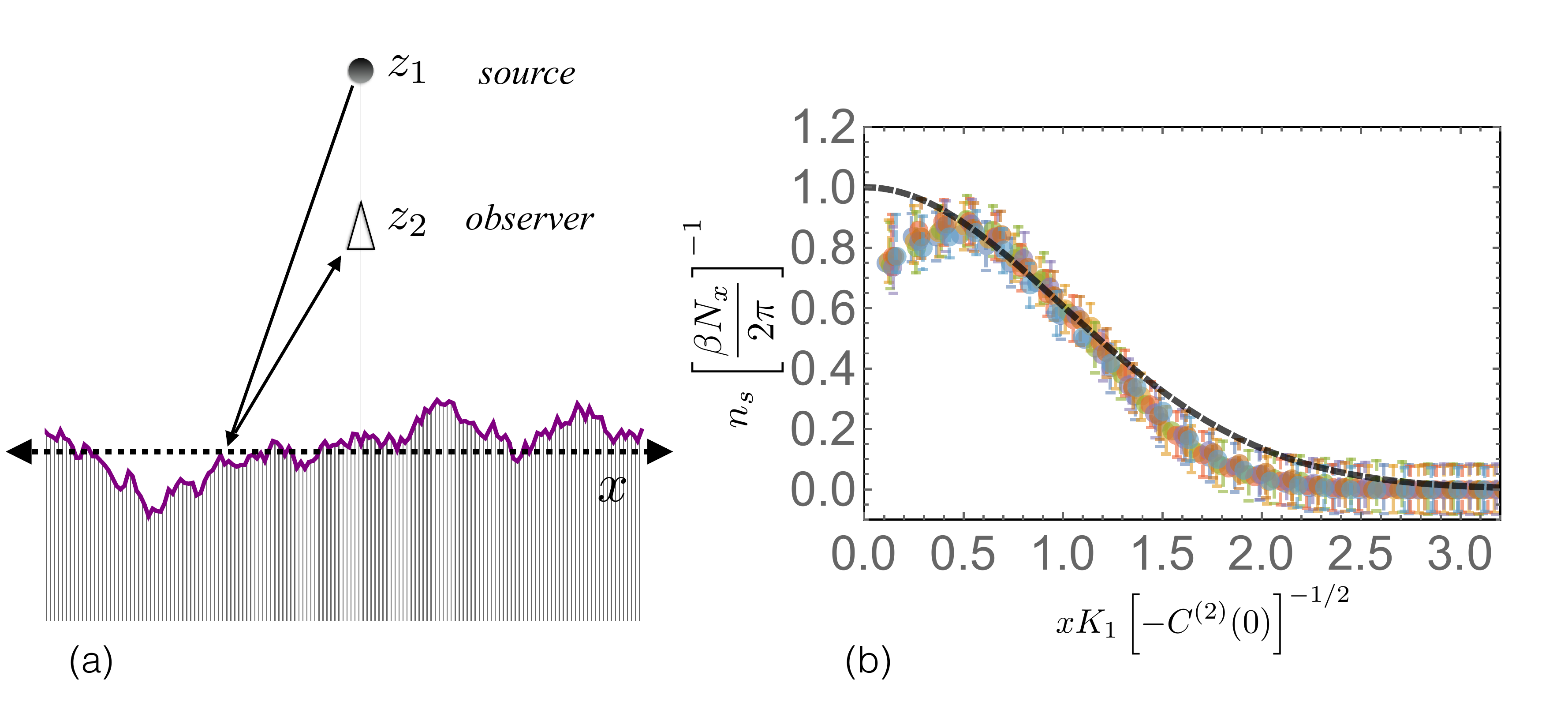}
\caption{(a) Schematic of the source, observer and a rough reflecting surface with the source and observer aligned above the origin of the $x$. Specular reflections become more and more improbable as $|x|$ becomes large. (b) Plot of the scaled density of specular points in the small angle approximation as a function of the distance from the common position of the source and observer along the $x-$axis. }
\label{fig_3}
\end{figure*}
where $C^{(2)}(0)=-\langle\big(\frac{df(x)}{dx}\big)^2\rangle$ and $C^{(4)}(0)=\langle\big(\frac{d^2f(x)}{dx^2}\big)^2\rangle$ are given in terms of the power spectrum $E(q)=\langle | f(q) |^2\rangle$ of $f(x)$ by,
\begin{subequations}
\begin{align}
\label{dauto}
C^{(2)}(0)&=-\int^{\infty}_{-\infty} q^2 E(q)dq < 0, \\
C^{(4)}(0)&=\int^{\infty}_{-\infty} q^4 E(q)dq. 
\end{align}
\end{subequations}
The probability distribution entering Eq. (\ref{pp1}), is thus given by,
\begin{eqnarray}
\label{pd}
P(g_1,g_2;x)=\frac{|x|}{2\pi \left[-C^{(2)}(0)C^{(4)}(0)\right]^{1/2}}\exp\left[\frac{g_1^{2}x^2}{2C^{(2)}(0)}-\frac{g^{2}_{2}}{2C^{(4)}(0)}\right].
\end{eqnarray}
Upon inserting Eq. (\ref{pd}) into Eq. (\ref{pp1}), we find the one-dimensional density of spectral points $n_s\equiv P_{spec}(x,K_1)$
for near normal angles of incidence and reflection as,
\begin{eqnarray}
\label{nx_long}
 n_s&=&\int^{\infty} _{-\infty} dg_{2}  \left[\frac{|K_1-g_2|}{|x|}\right]  \frac{|x|}{2\pi \left[-C^{(2)}(0)C^{(4)}(0)\right]^{1/2}}\exp\left[+\frac{K^{2}_1x^2}{2C^{(2)}(0)}-\frac{g^{2}_{2}}{2C^{(4)}(0)}\right]\nonumber\\\nonumber\\
&=& \exp\left[+\frac{K_1^{2}x^2}{2C^{(2)}(0)}\right] \frac{K_{1}}{ \left[-2\pi C^{(2)}(0)\right]^{1/2}}\left[\left(\frac{2}{\pi}\right)^{1/2}\beta^{-1}\exp\left[\frac{-\beta^2}{2}\right]
+erf\left(\frac{\beta}{\sqrt{2}}\right)\right]\nonumber\\
&=& N_s\exp\left[\frac{K_1^{2}x^2}{2C^{(2)}(0)}\right]\frac{K_1}{\left[-2\pi C^{(2)}(0)\right]^{1/2}}
\end{eqnarray}
where $N_s=\int^{\infty} _{-\infty}n_sdx$ and we define $\beta=K_{1}\left[C^{(4)}(0)\right]^{-1/2}$.
Using Monte-Carlo simulation of the fluctuating interfaces and membranes, we test the validity of Eq. (\ref{nx_long}) which predicts a Gaussian fall off of the density of specular points as a function of distance from the common location $(x=0)$ of the source and the observer. Figure 4.b depicts the results of the numerical simulations (see Sec. IV) represented by points and the prediction of the continuum theory in Eq. (\ref{nx_long}) which is given by the dashed curve. The small deviations from a Gaussian at small $x$ and large $x$ are likely due to higher order gradient couplings, neglected in our hydrodynamic approach.   
 

\subsection{Distribution of specular points in the large angle approximation:} 
We now consider a source and observer at more general coordinates $(x_{i},z_{i})$, still making the paraxial approximation ($z_j>> |x_j|$ and $z_{i}>>f(x)$) in Eq. (\ref{li}) but considering now the limit $|x_{i}|>>x$, i.e., source and observer reside far to the left and right of a specular point at $x$. Thus, we consider the configurations, in which the observation and incident angles measured form the normal and given by $\tan\theta_{i}=|x_i|/z_i<<1$  are finite and large compare to the angles $\tan^{-1}\left(\frac{|x|}{z_j}\right)$ that give rise to the highest density of specular points.
With these assumptions, vanishing of the first derivative of the total length $l=l_{1}+l_{2}$ defined in Eq. (\ref{li}) with respect to $x$ 
leads to the condition
\begin{eqnarray}
\label{ray_2}
\frac{ d f(x)}{dx}&=&-\frac{1}{2}\left[\frac{x_1}{z_1}+\frac{x_2}{z_2}\right]\equiv k_2,
\end{eqnarray}
where the source point and the observer reside at $S=(x_1,z_1)$ and $O=(x_2,z_2)$ respectively. Similar to the scheme we developed in the last section, we now define the auxiliary functions,
\begin{eqnarray}
\label{aux_II}
g_{1}(x)&\equiv&f'(x)=\frac{ d f(x)}{dx}\nonumber\\
 g_{2}(x)&\equiv&f''(x)=\frac{d^2 f(x)}{dx^2} 
\end{eqnarray}
where $g_1(x)=k_2$ is the condition for a specular reflection. To calculate the density of specular points, we need the joint probability distribution function $P[g_2,g_1;x]$ which can be calculated explicitly via functional integral methods (see Appendix B),
\begin{eqnarray}
\label{pd_3}
P\left(g_1,g_2;x\right)&=&\frac{\int \mathcal{D}f(y)\delta\left[f'(x)-g_1\right]\delta\left[f''(x)-g_2\right]e^{-F\left[f(y)\right]}}{\int \mathcal{D}f(y) e^{-F\left[f(y)\right]}}\nonumber\\ \nonumber\\ &=&\frac{1}{2\pi \left[-C^{(2)}(0)C^{(4)}(0)\right]^{1/2}}\exp\left[\frac{g_1^{2}}{2C^{(2)}(0)}-\frac{g^{2}_{2}}{2C^{(4)}(0)}\right],
\end{eqnarray}
which is independent of $x$ (provided $|x|<<|x_j|$) and where we have again assumed a quadratic free energy for interfaces and membranes such as Eq. (\ref{Fe_intro}) and (\ref{Fe_intro_mem}). The probability distribution for specular reflections is now,
\begin{eqnarray}
\label{pd_inf_I}
p_{spec}(x,k_2)dx&=&\int^{\infty} _{-\infty}dg_2 P(g_1,g_2;x)\big|\frac{dg_1}{dx}\big| dx\nonumber\\
&=&\int^{\infty} _{-\infty}dg_2 P(g_1,g_2;x)|g_2|dx,
\end{eqnarray}
which leads to an $x$-independent density of specular points,
\begin{eqnarray}
\label{pd_inf}
 n_{s}=\frac{1}{\pi}\left(\frac{C^{(4)}(0)}{- C^{(2)}(0)}\right)^{1/2}\exp\left[\frac{-k^2_{2}}{2C^{(2)}(0)}\right],
\end{eqnarray}
where 
\begin{eqnarray}
\label{zz}
C^{(4)}(0)=\lim_{y\rightarrow0}\langle f''(x)f''(x+y)\rangle.
\end{eqnarray}
Note that the density of specular points is proportional to $\left[C^{(4)}(0)\right]^{1/2}=\left[\langle\left(\partial^2_{x} f\right)^2\rangle\right]^{1/2}$, i.e. to the root mean square fluctuations in {\it curvature} of the reflecting curve. 
We can now use the fluctuating interface and membrane models introduced in Sec. II to study the statistical geometry of the specular points on a membrane in thermodynamic equilibrium. We express this density in terms of interface or membrane correlation length $\xi_I$, macroscopic size $L$ and microscopic dimension  $\ell$ of order the thickness. We focus on the scaling of density of the specular points at infinity, so that $k_2=\left[x_1/z_1+x_2/z_2\right]/2=0$. In this extreme paraxial limit, the density $n_{s}$, is characterized only by derivatives of the correlation functions $C(x)=\langle f(x)f(x+y)\rangle$, as in Eq. (\ref{pd_inf}). Upon using the relations in Eqs. (\ref{Co2_1}) -(\ref{Co2_3}) for the capillary-gravity interface, in the limit $\xi\gg\ell$ and $c=0$ we obtain,   
\begin{eqnarray}
\label{n_sp}
\lim _{ \xi\rightarrow 0} n_{s}=\frac{1}{\pi}\left(\frac{C^{(4)}(0)}{- C^{(2)}(0)}\right)^{1/2}\approx\frac{1}{\sqrt{3}\ell}
\end{eqnarray}
Using Eqs. (\ref{Ccapp})-(\ref{Ccapp-c}) for fluid membranes (or the equivalent model of fluctuating stretched two-dimensional polymer) yields,
\begin{eqnarray}
\label{n_sp_II}
\lim _{ \xi\rightarrow 0} n_{s}=\frac{1}{\pi}\left(\frac{C^{(4)}(0)}{- C^{(2)}(0)}\right)^{1/2}\approx\frac{1}{\sqrt{\ell\xi_{M}}}
\end{eqnarray}
Eqs. (\ref{n_sp}) and (\ref{n_sp_II}) demonstrate that the density of specular points diverges for both models as the microscopic cutoff, $\ell\rightarrow 0$. Although the specular density on a capillary-gravity wave interface is dominated by the short distance physics and is independent of the correlation length, the density of specular reflections for membranes, $n^{M}_{s}$ is controlled by the reciprocal of the $\it{geometric}$ $\it{mean}$ of the membrane correlation length and microscopic cutoff. Fluid membranes generate large number of specular reflections only in the limit of short correlation lengths $\xi_M > \ell$.

The densities of maxima, $n_{M}$ and minima, $n_{m}$ for a fluctuating random curve are closely related to the density of specular points in the extreme paraxial limit when both the observer and source reside at infinity. The specular point condition at infinity reduces to $f'(x)=k_2=0$ which of course is equivalent to being at a maximum or minimum. Since the average density of maxima and minima must be equal, they are related to the density of specular points calculated above by $n_{m}=n_{M}=n_{s}/2$. 
We can also investigate the ratio of $n_{0}$, the density of zero crossings to the density of maxima $n_{M}$ or minima $n_m$ for interfaces and membranes. 
With the aid of Eqs. (\ref{nxfC})-(\ref{nxfC_2}) and (\ref{n_sp})-(\ref{n_sp_II}) we find $n_{0}/n_{M}=n_{0}/n_m\approx\sqrt{\ell/\xi_M}\ll 1$ and $n_{0}/n_{m}=n_{0}/n_{M}\approx\sqrt{\ell/L}\ll 1$ for the capillary-gravity interface and fluid membrane model respectively. Thus the density of zero crossings is smaller than the density of maxima (or minima), which is plausible since for a highly fluctuating curve there can exist large number of maxima and minima between two consecutive zero crossings.

\begin{figure*}
\center\includegraphics[width=1.0\textwidth]{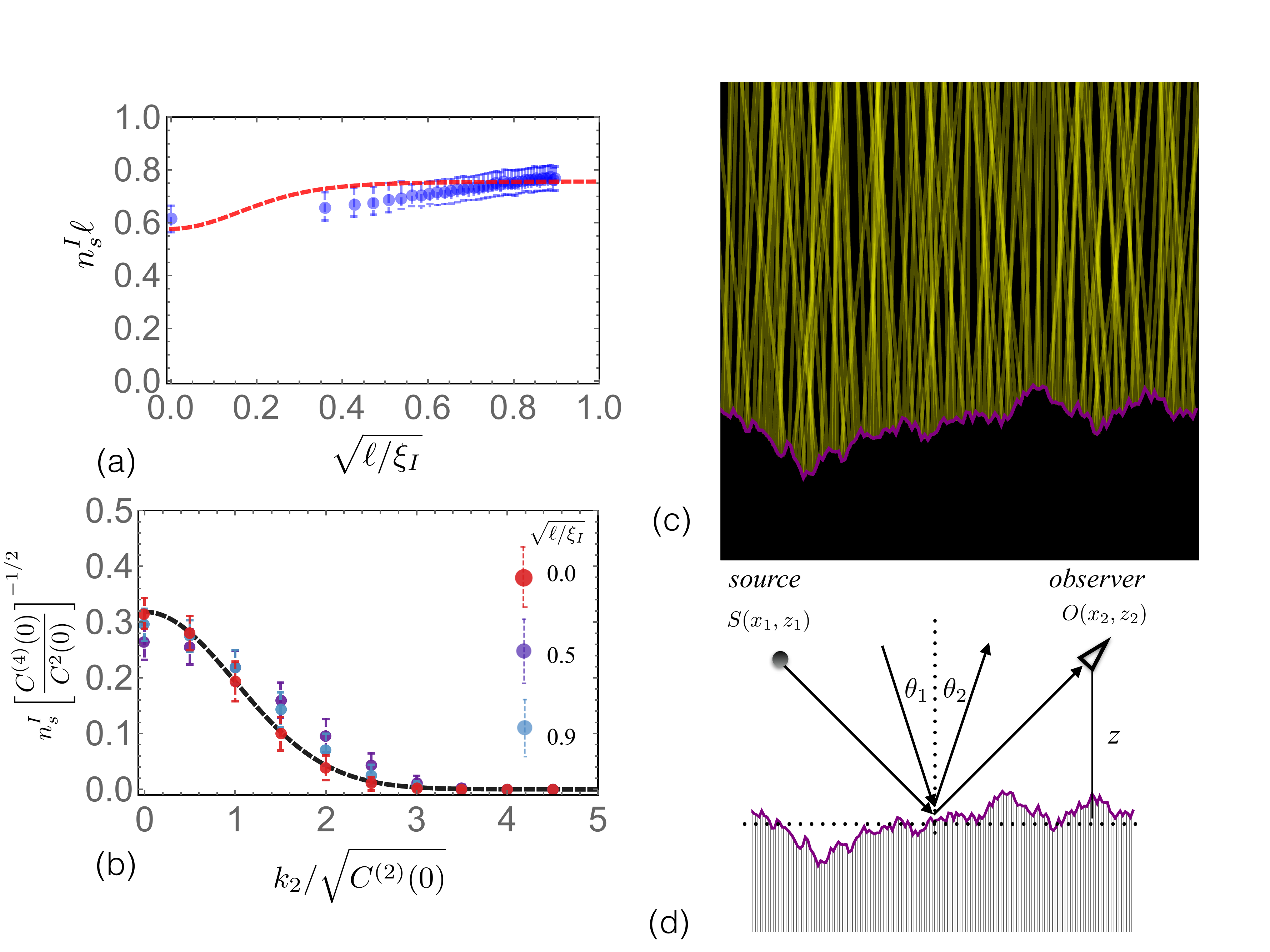}
\label{f2}
\caption{(a) the average density of specular points for a simulated $1$d surface with capillary-gravity waves, $c=0$ represented with points and the corresponding error bars. The dashed curve shows the analytical results, Eqs. (\ref{n_sp}) and (\ref{n_sp_II}), based on the paraxial approximation for the average density of specular points, $n_{s}$ scaled by the lattice constant, as a function of $\sqrt{\ell/\xi_I}$ for the 1D membrane with $L/\ell=120$ and $k_{B}T/(b\ell)=0.1$. (b) The dependence of $n_s$, the density of specular points as a function $k_{2}$ normalized by the average fluctuation amplitude for $\sqrt{\ell/\xi_I}=0,0.5,0.9$. The schematic in (c) illustrates a simulated reflecting membrane with$\sqrt{\ell/\xi_I}=0.36$ (only outgoing rays shown) with $k_2=0$ and the reflecting light rays of a light source at infinity. (d) The schematic of the source, observer and reflecting surface in large and small angle approximations}
\end{figure*}


\begin{figure*}
\center\includegraphics[width=1.0\textwidth]{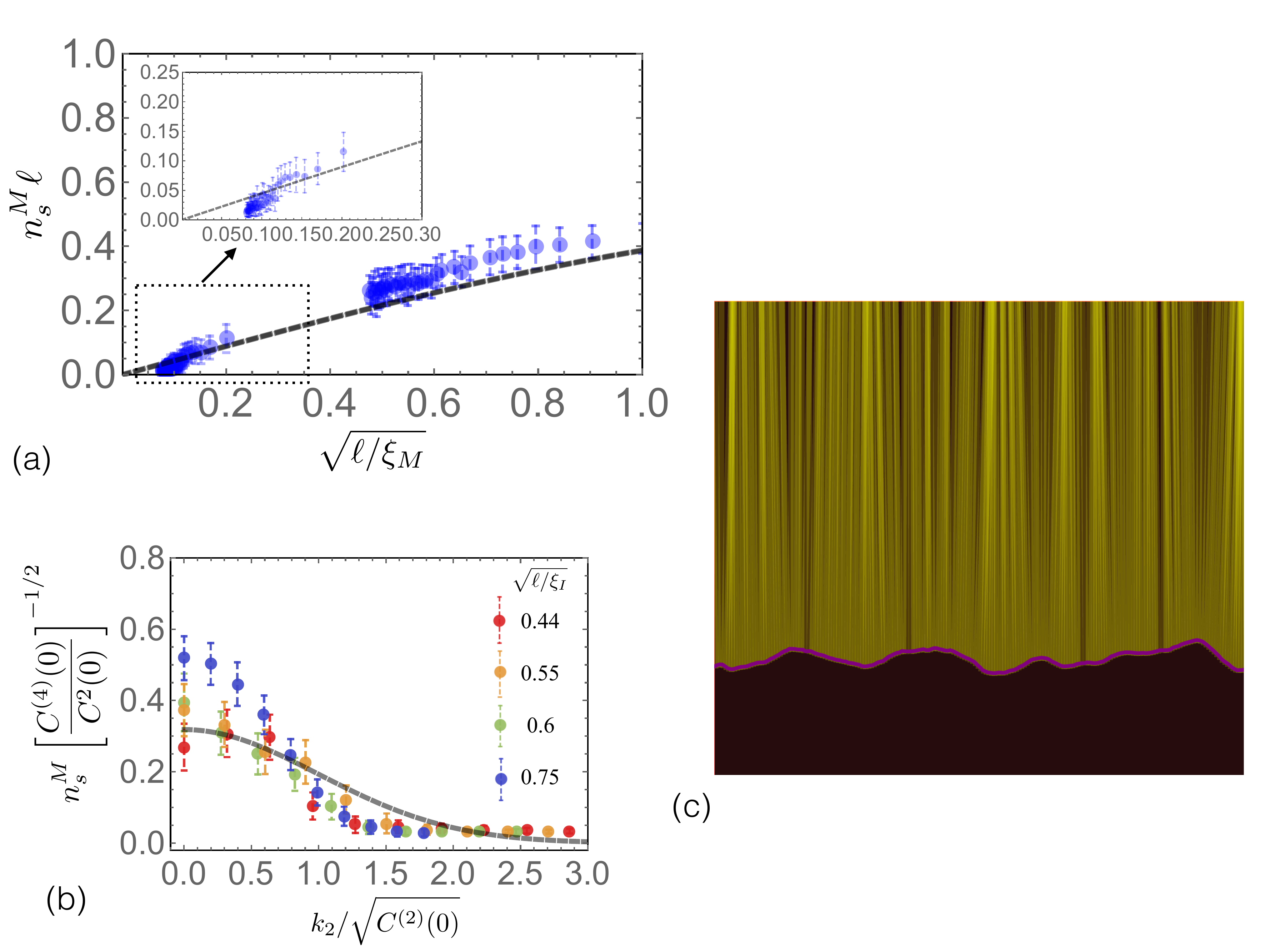}
\label{f5}
\caption{(a) the average density of specular points as derived analytically and from simulations of $1$d membrane model ($a=0$) with $N=300$ points. Points with error bars depict Monte-Carlo simulations. The dashed curve shows the analytical results based on the paraxial approximation for the average density of the scaled specular points, $n_{s}\ell$, as a function of $\sqrt{\ell/\xi}$ with $N=300$ and $k_{B}T/(b\ell)=0.1$ for $k_2=0$.  (b) the dependence of scaled $n_s$ as a function $k_{2}/\sqrt{C^{(2)}(0)}$, which characterizes the distance of the observable from the reflecting plane presented in the plot (b) for different values of $\sqrt{\ell/\xi_{M}}=0.44,0.55,0.6,0.75$. (c) a simulated membrane with $\sqrt{\ell/\xi_{M}}=0.1$ and $k_2=0$. The gold lines above the interface show the reflected light rays of a light source at infinity (only outgoing rays shown). The bands of enhanced density of these rays suggest that caustics will from sufficiently far from the interface}
\end{figure*}


\subsection{A Monte Carlo simulations of fluctuating one dimensional chain model}

To explore the validity of expressions for the density of zero crossings and specular points given for interfaces and membranes in terms of $C(x)=\langle f(y+x)f(x)\rangle$ and its derivatives at the origin ($C(0)$, $C^{(2)}(0)$ and $C^{(4)}(0)$), we carry out Monte-Carlo simulations. We simulate a fluctuating $1$-d surface in two dimensions described by the discretized free energy,
\begin{eqnarray}
\label{fdis}
F_{dis}=\sum_{j=1}^{N} \left[ \frac{a'}{4}\left(f_{j}^2+f_{j+1}^2\right)+\frac{b'}{2}\left(f_{j}-f_{j+1}\right)^2+\frac{c'}{2}\left(2f_{j}-f_{j-1}-f_{j+1}\right)^2 \right],
\end{eqnarray}
where in the continuum limit our simulation parameters $a',b'$ and $c'$ are related to the parameters appearing in Eq. (\ref{Fe}) by $a'=a\ell$, $b'=b/\ell$ and $c'=c/\ell^3$. 
The height function $f(x)$ is defined in the interval $(0,L)$ where $L$ is the system size with periodic boundary conditions. It is discretized by considering the height at two neighboring points, $f_{j}$ and $f_{j+1}$, where $f_{j}=f(x=x_j)$ is defined in the small interval $(x_j,x_{j+1})=(x_{j}+\ell)$, $j=1,...,N$ with $N=L/\ell$. The continuum energy expression in Eq. (\ref{Fe}) can be recovered via taking the limit $\ell/L\rightarrow 0$ in eq. (\ref{fdis}).

Monte Carlo simulations are performed on this ``chain model" at temperature $k_{B}T$ (in units of $b\ell$) with the following move: a random vertex $i$ residing at $(x_{i},y_{i}=f(x_i))$ is chosen and displaced to $(x_{i}+\delta x,y_{i}+\delta y)$ where $\delta x$ and $\delta y$ are small independent random real numbers. Moves are accepted or rejected based on the Metropolis criterion, using the Boltzmann factor $\exp\left[-F_{dis}/k_B T\right]$ associated with the free energy in Eq. (\ref{fdis}). By tracking the time-dependance of correlation functions, we found that system with $N=150$ equilibrates after $\approx 10^6$ MC steps. One MC step is defined as $N$ vertex moves.
Thermodynamic quantities like the average number of zero crossings or specular points, are measured after equilibration has been reached.   
We performed the simulations by starting from a configuration where all $N$ points reside on a straight line, equally spaced with separation $\ell$.
The results of the Monte Carlo simulations for the average density of zero crossings, $n_{0}$, of the
fluctuating membrane illustrated in Fig. 2 and Fig. 3. For the capillary-gravity interface model we take  $c'=0$ in Eq. (\ref{fdis}), while for the membrane or polymer-like model, we take $a'=0$. The zero crossings are shown as a function of the ratio of the lattice spacing to the correlation length $\sqrt{\ell/\xi}$ in Figs. 2 and 3 with the corresponding error bars. The dashed line represents the analytical result for the density of zero crossings in eq. (\ref{nxfC}) with the correlation function calculated based on the free energy with $c=0$, given by Eqs. (\ref{Co2_1})-(\ref{Co2_3}). In fig.2b we confirm the Gaussian fall-off of the expected number of $f(x)=y$ passages, $n(y)$, as a function of the scaled specific height $y/\sqrt{C(0)}$, for $\sqrt{\ell/\xi}=0,0.5,0.9$. A similar comparison has been done for the density of zero crossings for the membrane model in Fig 3b. 

Fig.4b shows the dependence of the average density of specular points on $x$ in the small angle approximation using the Monte-Carlo simulations (points) and the predictions of the continuum theory (dashed line) given in Eq. (28). It confirms the Gaussian fall off of the average density of specular points in agreement with Eq. (28) for the thermally excited interfaces.  
Next we simulate the density of specular reflection points from a thermally excited interface dressed with the capillary-gravity waves ($c=0$) as a function of the correlation length, $\xi$ and the distance of the observer from the $1d$ membrane in terms of $k_2=\frac{-1}{2}\left[\frac{x_1}{z_1}+\frac{x_2}{z_2}\right]$ in the extreme paraxial limit ($|x_j|\ll z_j$). Figure 5a and 6a illustrate the density of specular points at infinity when $k_2=0$ as function of $\sqrt{\ell/\xi}$, where $\xi$ is the relevant interface or membrane correlation length. We show the dependence of the density if specular points on the distance of the observer from the reflecting line which is encoded in $k_2$ and described by Eq. (\ref{pd_inf}). The density of specular points was evaluated in the simulation by measuring the slope of the fluctuating curves and computing the expected number of times the local slope satisfies the condition for specular reflection $\frac{df}{dx}=k_2$ given by Eq. (\ref{ray_2}). 
The density of specular points reaches a constant value for large $\xi$. 
The schematic presented in Fig.5c shows a simulated reflecting interface and the reflected light rays of a light source at infinity. The incoming light rays parallel to $z$ axis and has not been shown. The number of reflecting light rays reaching the point of observation determines the number of specular points   
Similarly Fig 6a-c, depicts the result of the simulations for the fluctuating curve described via $a=0$ which mimics the behavior of the fluctuating membrane or polymer under tensile stresses characterized by the coefficient $b$ and the bending modulus $c$. Fig.6a shows the expected number of specular points as a function of $\sqrt{\ell/\xi_M}$ for $k_2=0$ (observer at infinity) governed by eq. (\ref{Ccappc}) and Fig.6b illustrates the $k_2$ dependence for different values of $\sqrt{\ell/\xi_M}$. We also show the dependence of the density of specular points on the distance from the reflection center $(x=0)$ on the reflecting surface has been shown in Fig4.b for both membrane and interface models described by $c=0$ and $a=0$. The result of the simulations is compared to the analytical calculation presented in eq. (\ref{nx_long}) in the small angle approximation.    


\section{conclusion}
To explore the statistical geometry of specular reflections from fluctuating surfaces in thermal equilibrium, we developed a mapping which connects configurations of thermalized surfaces described by the relevant elastic constants to the statistics of zero crossings and specular reflection points. The density of the specular points is proportional to the root mean square curvature of the interface or membrane.  
This study can be generalized to the statistics of fold, cusp and higher order caustics to construct a more complex mapping between the geometry, mechanics and thermodynamic of a surface and the statistical geometry of reflected caustics patterns in higher dimensions. More generally, this study suggests new avenues for surface characterization based on the distribution of specular patterns forms in a plane with dimensions lower that the dimension of the embedding space. This framework can be used to infer material properties such as surface tension and correlation lengths of thermalized interfaces and membranes, by studying the lower dimensional pattern of reflections.    

Upon zooming in the blurred reflected pattern of coherent light even at smaller length scales, one expects to observe the fine structure of the diffraction pattern associated with the optical wave field, consisting of a complex pattern of dislocation lines, which are disruptions in the reflected wavefront \cite{wave_1,wave_2,wave_3,wave_4}. Future studies would be needed to reveal how the geometry and statistical mechanics of the fluctuating interfaces control the fine structure of the diffraction pattern. 

\section{Acknowledgments}
It is a pleasure to acknowledge stimulating interactions with Massimo Cencini, Daniel A. Beller and Yoav Lahini. We are grateful to Michael Berry for comments on the manuscript. This work was supported by the National Science Foundation, through grants DMR-1608501 and via the Harvard Materials Science Research and Engineering Center via grant  DMR-1435999. We would like to dedicate this article to the memory of Pierre C. Hohenberg, whose pioneering work on dynamic critical phenomena with B.I. Halperin and others had a profound influence early in the career of us.

\section{Appendices}
\appendix

\section{Correlation function calculation}

Here we provide details of the correlation function calculations reported in section II. A for the fluctuating interfaces or membranes in $1+1$ dimensions. From Eq. (\ref{Fe}) we obtain, 
\begin{eqnarray}
\label{en}
C(y)=\langle f(x_{0})f(x_{0}+y)\rangle=\frac{1}{2\pi}\int^{+\infty}_{-\infty}\frac{dq e^{iqy}}{a+b q^2+ c q^4}
\end{eqnarray}
Upon considering the complex $q$-plane, the denominator defines four poles at:
\begin{eqnarray}
\label{en}
q&=&\pm i \left[\frac{b\mp\sqrt{\delta}}{2c}\right]^{1/2}\nonumber\\
\delta&=&b^{2}-4ac
\end{eqnarray}  
We assume that $\delta>0$. Upon defining an interfacial capillary-gravity correlation length by $\xi_I\equiv\sqrt{b/a}$ and the ultraviolet cutoff induced by the $c$-term as $\ell\equiv\sqrt{c/b}$, this condition simply means that $\xi_I>2\ell$. Because all poles reside on the imaginary axis, we also have $b>\sqrt{\delta}$. By completing the contour in the upper half plane, which encloses the poles at $q= i \left[\frac{b\pm\sqrt{\delta}}{2c}\right]^{1/2}$, we find,
\begin{eqnarray}
\label{Cc_app}
C(y)=\frac{1}{2\pi}\int^{+\infty}_{-\infty}\frac{dq e^{iqy}}{a+b q^2+ c q^4}&=& i \bigg(\frac{e^{iqy}}{2bq+4cq^3}\bigg)_{q=i \left[\frac{b+\sqrt{\delta}}{2c}\right]^{1/2}}+ i \bigg(\frac{e^{iqy}}{2bq+4cq^3}\bigg)_{q=i \left[\frac{b-\sqrt{\delta}}{2c}\right]^{1/2}}\nonumber\\
&=&\frac{1}{\sqrt{2 \delta}}\left[\frac{e^{-\sqrt{(\frac{b-\sqrt{\delta}}{2c})}|y|}}{\sqrt{\frac{b-\sqrt{\delta}}{c}}}-\frac{e^{-\sqrt{(\frac{b+\sqrt{\delta}}{2c})}|y|}}{\sqrt{\frac{b+\sqrt{\delta}}{c}}}\right]
\end{eqnarray}
In a similar fashion we find that the second derivative of the correlation function is given by,
\begin{eqnarray}
\label{Ccd_app}
C^{(2)}(y)&=&\frac{-1}{2\pi}\int^{+\infty}_{-\infty}\frac{q^2 dq e^{iqy}}{a+b q^2+ c q^4}\nonumber\\
&=&\frac{1}{2\sqrt{2 \delta}}\left[\sqrt{\frac{b-\sqrt{\delta}}{c}}e^{-\sqrt{(\frac{b-\sqrt{\delta}}{2c})}|y|}-\sqrt{\frac{b+\sqrt{\delta}}{c}}e^{-\sqrt{(\frac{b+\sqrt{\delta}}{2c})}|y|}\right]
\end{eqnarray}
Note that $C^{(2)}(0)$ diverges as $c\rightarrow0$, reflecting the sensitivity of this quantity to the effective ultraviolet cutoff $\ell=\sqrt{c/b}$ as $\ell\rightarrow0$.

\section{Functional integrals for the probability density of a fluctuating membrane}
Here we calculate, via functional integral methods, the normalized probability of a function $f(y)$ to have a specific value and slope at position $x$ given by $f(x)=g, f'(x)=g_1$. The probability density for these quantities with a fluctuating interface described by our simple quadratic free energy functional $F_{I}\left[f(y)\right]$, (we use the interface free energy Eq. (\ref{Fe_ab}) for concreteness, but identical manipulation apply for the membrane free energy Eq. (\ref{Fe_bc})) is given by, 
 \begin{eqnarray}
\label{FI}
p\left[f(x)=g,f'(x)=g_1;x\right]&=&\frac{\int \mathcal{D}f(y)\delta\left[f(x)-g\right]\delta\left[f'(x)-g_1\right]e^{-F_I\left[f(y)\right]}}{\int \mathcal{D}f(y) e^{-F_I\left[f(y)\right]}}\nonumber\\
&=&
\frac{\int \mathcal{D}f(y)\delta\left[f(x)-g\right]\delta\left[f'(x)-g_1\right]e^{\frac{-1}{2}\int dy\left[a f(y)^{2}+b\left(\frac{df(y)}{dy}\right)^{2}\right]}}{\int \mathcal{D}f(y) e^{-F_I\left[f(y)\right]}}\nonumber\\&=&
\int ds\int dt e^{-2\pi i sg}e^{-2\pi i t g_1}\langle e^{2\pi i s f(x)+2\pi i t f'(x) }\rangle\nonumber\\&=&
\int ds\int dt e^{-2\pi i sg}e^{-2\pi i t g_1}e^{-2\pi^2 \langle f^2(x)\rangle s^2 -2\pi^2 \langle f'^2(x)\rangle t^2-4\pi^2 \langle f(x) f'(x)\rangle st  }\nonumber\\&=&
\int ds\int dt e^{-2\pi i sg}e^{-2\pi i t g_1}e^{-2\pi^2 C(0) s^2+2\pi^2 C^{(2)}(0) t^2  }\nonumber\\
\end{eqnarray}
where $\langle .\rangle$ represents a thermal average, and
$\left\langle f(x)f'(x)\right\rangle=\frac{1}{2}\left\langle\frac{d}{dx}f^{2}(x)\right\rangle=0$ with our periodic boundary conditions. We also have $C(0)=\langle f^{2}(x)\rangle$ and $C^{(2)}(x)=-\left\langle\left(\frac{df}{dx}\right)^{2}\right\rangle$.
Here we used the identity, valid for any Gaussian probability distribution, $\langle e^{h(x)}\rangle=e^{\frac{1}{2}\langle{h^{2}(x)}\rangle}$ and the integral representation of the delta function,
\begin{eqnarray}
\label{id}
\delta\left[f(x)-g\right]=\int^{\infty}_{-\infty} ds e^{2\pi i \left[f(x)-g\right]s} .
\end{eqnarray}
This probability density for a fluctuating membrane is normalized by construction, $\int^{\infty}_{-\infty} p\left[g,g_1\right]dg_1,dg_2=1$.
By the aid of standard Gaussian integrals we find from Eq. (\ref{FI}) the probability density, $p\left[f(x)=g, f'(x)=g_1;x\right]$ is in fact independent of $x$,
\begin{eqnarray}
\label{FII}
p\left[f(x)=g,f'(x)=g_1;x\right]=\frac{1}{2\pi \left[-C(0)C^{(2)}(0)\right]^{1/2}}\exp\left[-\frac{g^{2}}{2C(0)}+\frac{g^{2}_{1}}{2C^{(2)}(0)}\right].
\end{eqnarray}
Similarly, one can calculate the normalized probability of a function $f(y)$ to have a given slope and curvature described by $f'=g_1, f''=g_2$ at the position $x$,
\begin{eqnarray}
\label{FIII}
p\left[f'(x)=g_1,f(x)''=g_2;x\right]&=&\frac{\int \mathcal{D}f(y)\delta\left[f'(x)-g_1\right]\delta\left[f''(x)-g_2\right]e^{-F\left[f(y)\right]}}{\int \mathcal{D}f(y) e^{-F\left[f(y)\right]}}\nonumber\\
&=&
\int ds\int dt e^{-2\pi i sg_1}e^{-2\pi i t g_2}\langle e^{2\pi i s f'(x)+2\pi i t f''(x) }\rangle\nonumber\\&=&
\int ds\int dt e^{-2\pi i sg_1}e^{-2\pi i t g_2}e^{2\pi^2 C^{(2)}(0) s^2-2\pi^2 C^{(4)}(0) t^2  }\nonumber\\&=&
\frac{1}{2\pi \left[-C^{(2)}(0)C^{(4)}(0)\right]^{1/2}}\exp\left[\frac{g_1^{2}}{2C^{(2)}(0)}-\frac{g_2^{2}}{2C^{(4)}(0)}\right]
\end{eqnarray}
where $C^{(4)}(0)=\lim_{y\rightarrow0}\left\langle f''(x+y)f''(x)\right\rangle=\left\langle\left(\frac{d^2 f}{dx^2}\right)^2\right\rangle$. In a similar fashion, we can find the probability density of a stochastic function with fixed higher order derivatives, at position $x$, with $\frac{d^n f(x)}{dx^n}=g_n$.

\end{document}